\documentclass[12pt]{iopart}
\usepackage{graphicx}

\begin{document}

\title[Decoherence suppression of a...]{Decoherence suppression of a dissipative qubit by non-Markovian effect}
\author{Qing-Jun Tong$^{1}$, Jun-Hong An$^{1,2}$\footnote{%
Email: anjhong@lzu.edu.cn}, Hong-Gang Luo$^{1,3}$ and C. H. Oh$^{2}$}
\address{$^{1}$ Center for Interdisciplinary Studies, Lanzhou University, Lanzhou 730000,
China}
\address{$^{2}$ Department of Physics, National University of Singapore, 3 Science Drive 2, Singapore 117543, Singapore}
\address{$^{3}$ Key Laboratory for
Magnetism and Magnetic materials of the Ministry of Education,
Lanzhou University, Lanzhou 730000, China}

\begin{abstract}
We evaluate exactly the non-Markovian effect on the decoherence
dynamics of a qubit interacting with a dissipative vacuum reservoir
and find that the coherence of the qubit can be partially trapped in
the steady state when the memory effect of the reservoir is
considered. Our analysis shows that it is the formation of a bound
state between the qubit and its reservoir that results in this
residual coherence in the steady state under the non-Markovian
dynamics. A physical condition for the decoherence suppression is
given explicitly. Our results suggest a potential way to decoherence
control by modifying the system-reservoir interaction and the
spectrum of the reservoir to the non-Markovian regime in the
scenario of reservoir engineering.
\end{abstract}

\maketitle



\section{Introduction}
Any realistic quantum system inevitably interacts with its
surrounding environment, which leads to the loss of coherence, or
decoherence, of the quantum system \cite{Breuer02}. The decoherence
of quantum bit (qubit) is deemed as a main obstacle to the
realization of quantum computation and quantum information
processing \cite{Nielsen00}. Understanding and suppressing the
decoherence are therefore a major issue in quantum information
science. For a Markovian environment, it is well known that the
coherence of a qubit experiences an exponential decrease
\cite{Breuer02}. To beat this unwanted degradation, many controlling
strategies, passive or active, have been proposed
\cite{decohControl1,decohControl2,decohControl3,decohControl4,decohControl5}.

In recent years much attention has been paid to the non-Markovian
effect on the decoherence dynamics of open quantum system
\cite{Garraway97, Maniscalco2006, Pillo08, Koch08, Breuer08}. The
significance of the non-Markovian dynamics in the study of open
quantum system is twofold. i) It is of fundamental interest to
extend the well-developed methods and concepts of Markonian dynamics
to non-Markovian case \cite{Breuer02, Gardiner} for the open quantum
system in its own right. ii) There are many new physical situations
in which the Markovian assumption usually used is not fulfilled and
thus the non-Markovian dynamics has to be introduced. In particular,
many experimental results have evidenced the existence of the
non-Markovian effect \cite{Dubin07,Koppens07,Mogilevtsev08}, which
indicates that one can now approach the non-Markovian regime via
tuning the relevant parameters of the system and the reservoir. The
non-Markovian effect means that the environment, when its state is
changed due to the interaction with the quantum system, in turn,
exerts its dynamical influence back on the system. Consequently one
can expect decoherence dynamics of the quantum system could exhibit
a dramatic deviation from the exponential decaying behavior. In
2005, DiVincenzo and Loss studied the decoherence dynamics of the
spin-boson model for the Ohmic heat bath in the weak-coupling limit.
They used the Born approximation and found that the coherence
dynamics has a power-law behavior at long-time scale \cite{Loss05},
which greatly prolongs the coherence time of the quantum system.
Such power-law behavior suggests that the non-Markovian effect may
play a constructive role in suppressing decoherence of the system.
Nevertheless, in many cases the finite extension of the coherence
time of the system is not sufficient for the quantum information
processing, a question arises whether the coherence of the system
can be preserved in the long-time limit, even partially.
Theoretically, the answer is positive if the environment has a
nontrivial structure. It has been shown that some residual coherence
can be preserved in the long-time steady state when the environment
is a periodic band gap material
\cite{Yablonovitch87,Sajeev94,Lambropoulos00,Bellomo08} or leaky
cavity \cite{Scala08}. It is stressed that the residual coherence is
due to the confined structured environment. A natural question is:
Whether the coherence of the system can be dynamically preserved or
not by the non-Markovian effect if the environment has no any
special structure, e.g., a vacuum reservoir?

In this paper, we study the exact decoherence dynamics of a qubit
interacting with a vacuum reservoir and examine the possibility of
decoherence suppression using the non-Markovian effect. The main aim
of this work is to analyze if and how the coherence present in the
initial state can be trapped with a noticeable fraction in the
steady state even when the environment is consisted of a vacuum
reservoir with trivial structure. We show that the non-Markovian
effect manifests its action on the qubit not only in the transient
dynamical process, but also in the asymptotical behavior. Our
analysis shows that the physical mechanism behind this dynamical
suppression to decoherence is the formation of a bound state between
the qubit and the reservoir. The no-decaying character of the bound
state leads to the inhibition of the decoherence and the residual
coherence trapped in the steady state. A similar vacuum induced
coherence trapping in the continuous variable system has been
reported in \cite{An08,An09}. Such coherence trapping phenomenon
provides an alternative way to suppress decoherence. This could be
realized by controlling and modifying the system-reservoir
interaction and the properties of the reservoir
\cite{Yablonovitch87} by the recently developed reservoir
engineering technique \cite{Wineland001,Wineland002,Diehl08}.

Our paper is organized as follows. In Sec. \ref{model}, we introduce
the model of a qubit interacting with a vacuum reservoir and derive
the exact master equation. In Sec. \ref{quant}, two quantities, i.e.
purity and decoherence factor, to characterize the decoherence
dynamics are introduced. In Sec. \ref{numa} we give the numerical
study for the time evolution of decay rate, purity and decoherence
factor in terms of coupling constant and cutoff frequency and the
physical mechanism of the dynamical decoherence suppression.
Finally, discussions and summary are given in Sec. \ref{dsumma}

\section{The model and exact dynamics of the qubit}\label{model}

We consider a qubit interacting with a reservoir which is consisted
of a quantized radiation field. The Hamiltonian of the total system
is given by
\begin{equation}
H=\omega _{0}\sigma _{+}\sigma _{-}+\sum_{k}\omega _{k}a_{k}^{\dag
}a_{k}+\sum_{k}(g_{k}\sigma _{+}a_{k}+h.c.),  \label{t1}
\end{equation}%
where $\sigma _{\pm }$ and $\omega _{0}$ are the inversion operators
and transition frequency of the qubit, $a_{k}^{\dag }$ and $a_{k}$
are the creation and annihilation operators of the $k$-th mode with
frequency $\omega _{k}$ of the radiation field. The coupling
strength between the qubit and the radiation field has the form
\cite{Breuer02}
\begin{equation}
g_{k}=-i\sqrt{\frac{\omega _{k}}{2\varepsilon
_{0}V}}\mathbf{\hat{e}} _{k}\cdot \mathbf{d},  \label{t19}
\end{equation}
where $\mathbf{\hat{e}}_{k}$ and $V$ are the unit polarization
vector and the normalization volume of the radiation field,
$\mathbf{d}$ is the dipole moment of the qubit, and $\varepsilon
_{0}$ is the free space permittivity. Throughout this paper we
assume $\hbar =1$. This model has been well studied under the
Born-Markovian approximation in quantum optics \cite{Breuer02}.
However, what is the physical condition under which such
approximation is applicable and how the non-Markovian effect affects
the decoherence dynamics in the different parameter regimes have not
been quantitatively investigated.

If there is no correlation between the qubit and the radiation field
initially, then the initial state of the whole system can be
factorized into a product of the states of qubit and the field. If
the initial state is $\left\vert \Psi (0)\right\rangle =\left\vert
+,\{0_{k}\}\right\rangle $, with $\left\vert +\right\rangle $ and
$|\{0_k\}\rangle$, respectively, denoting the exited state of the
qubit and the vacuum state of the radiation field, then governed by
the Hamiltonian (\ref{t1}),
the state will evolve to the following form%
\begin{equation}
\left\vert \Psi (t)\right\rangle =b_{0}(t)\left\vert
+,\{0_{k}\}\right\rangle +\sum_{k}b_{k}(t)\left\vert
-,1_k\right\rangle ,\label{am}
\end{equation}%
where $\left\vert 1_{k}\right\rangle $ is the field state
containing one photon only in the $k$-th mode. From the Schr\"{o}%
dinger equation, we can get the time evolution of the probability
amplitudes in Eq. (\ref{am}). On substituting the formal solution
of $b_k(t)$ into the equation of motion satisfied by $b_0(t)$, we obtain%
\begin{equation}
\dot{b}_{0}(t)+i\omega _{0}b_{0}(t)+\int_{0}^{t}b_{0}(\tau )f(t-\tau
)d\tau =0,  \label{t4}
\end{equation}%
where the kernel function is $f(x)=\sum_{k=0}^{\infty }\left\vert
g_{k}\right\vert ^{2}e^ {-i\omega _{k}x}$. The integro-differential
equation (\ref{t4}) renders the dynamics of the qubit non-Markovian,
with the memory effect of the reservoir registered in the
time-nonlocal kernel function $f(x)$. In the continuous limit of the
environment frequency, one can verify from the coupling strength
(\ref{t19}) that the kernel function has the form
\begin{equation}
f(x)=\int_{0}^{\infty }J(\omega )e^{-i\omega x}d\omega , \label{t15}
\end{equation}%
where $J(\omega )=\eta \frac{\omega
^{3}}{\omega_0^2}e^{\frac{-\omega }{\omega _{c}}}$
is called the spectral density with the coupling constant $%
\eta \equiv \omega_0^2\frac{\int \left\vert
\mathbf{\hat{e}}_{k}\cdot \mathbf{d}\right\vert ^{2}d\Omega }{(2\pi
c)^{3}2\varepsilon _{0}}$. Here the $\omega_0^2$ in $J(\omega_0)$ is
introduced to make $\eta$ dimensionless. To eliminate the infinity
in frequency integration, we have introduced the cutoff frequency
$\omega _{c}$. Physically, the introducing of the cutoff frequency
means that not all of the infinite modes of the reservoir contribute
to the interaction with the qubit, and one always expects the
spectral density going to zero for the modes with frequencies higher
than certain characteristic frequency. It is just this
characteristic frequency which determines the specific behavior and
the properties of the reservoir. One can see that in our model, the
spectral density has a super-Ohmic form \cite{Leggett87}.

From the time evolution of Eq. (\ref{am}) and the fact that the
ground state $|-\rangle$ of the qubit is immune to the radiation
field, one can get the time evolution of any given initial state of
the system readily. For a general state of the qubit described by
\begin{eqnarray*}
\rho _{tot}(0) &=&(\rho _{11}\left\vert +\right\rangle \left\langle
+\right\vert +\rho _{12}\left\vert +\right\rangle \left\langle
-\right\vert
+\rho _{21}\left\vert -\right\rangle \left\langle +\right\vert \\
&&+\rho _{22}\left\vert -\right\rangle \left\langle -\right\vert
)\otimes \left\vert \{0\}_{k}\right\rangle \left\langle
\{0\}_{k}\right\vert,
\end{eqnarray*}%
the time evolution of the total system can be calculated explicitly.
In fact, what is needed is the reduced density matrix of the qubit,
which is obtained by tracing
over the reservoir variables%
\begin{equation}
\rho (t)=\left(
\begin{array}{cc}
\rho _{11}\left\vert b_{0}(t)\right\vert ^{2} & \rho _{12}b_{0}(t) \\
\rho _{21}b_{0}^{\ast }(t) & 1-\rho _{11}\left\vert b_{0}(t)\right\vert ^{2}%
\end{array}%
\right) .  \label{t5}
\end{equation}%
Differentiating Eq. (\ref{t5}) with respect to time, we arrive at
the equation of motion of the reduced density matrix
\begin{eqnarray}
\dot{\rho}(t) &=&-i\frac{\Omega (t)}{2}[\sigma _{+}\sigma _{-},\rho (t)]+%
\frac{\gamma (t)}{2}[2\sigma _{-}\rho (t)\sigma _{+}  \nonumber \\
&&-\sigma _{+}\sigma _{-}\rho (t)-\rho (t)\sigma _{+}\sigma _{-}],
\label{t6}
\end{eqnarray}
where $\Omega (t)=-2Im[\frac{\dot{b}_{0}(t)}{b_{0}(t)}]$ and $\gamma
(t)=-2Re[\frac{\dot{b}_{0}(t)}{b_{0}(t)}]$. $\Omega (t)$ plays the
role of time-dependent shifted frequency and $\gamma (t)$ that of
time-dependent decay rate \cite{Breuer02}. It is worth mentioning
that in the derivation of the master equation (\ref{t6}), we have
not resorted to the Born-Markovian approximation, that is, Eq.
(\ref{t6}) is the exact master equation of the qubit system.
Compared with the master equation derived in Ref. \cite{Breuer02}
under the condition that the initial state of the qubit is pure, our
derivation shows that Eq. (\ref{t6}) can describe the dynamics not
only for the initial pure state, but also for any mixed state of the
qubit.

It is interesting to note that one can reproduce the conventional
Markovian one from our exact non-Markovian master equation under
certain approximations. By redefining the probability amplitude as
$b_{0}(\tau )=b_{0}^{\prime }(\tau )e^{-i\omega _{0}\tau }$, one can
recast Eq.~(\ref{t4} ) into
\begin{equation}
\dot{b}_{0}^{\prime }(t)+\int_{0}^{\infty }d\omega J(\omega
)\int_{0}^{t}d\tau e^{i(\omega _{0}-\omega )(t-\tau )}b_{0}^{\prime
}\left( \tau \right) =0. \label{t13}
\end{equation}%
Then, we take the Markovian approximation $b_{0}^{\prime }\left(
\tau \right) \cong b_{0}^{\prime }(t)$, namely, approximately taking
the dynamical variable to the one that depends only on the present
time so that any memory effect regarding the earlier time is
ignored. The Markovian approximation is mainly based on the physical
assumption that the correlation time of the reservoir is very small
compared with the typical time scale of system evolution. Also under
this assumption we can extend the upper limit of the $\tau $
integration in Eqs.~(\ref{t13}) to infinity and use the equality
\begin{equation}
\lim_{t\rightarrow \infty }\int_{0}^{t}d\tau e^{\pm i(\omega
_{0}-\omega )(t-\tau )}=\pi \delta (\omega -\omega _{0})\mp
iP\Big(\frac{1}{\omega -\omega _{0}}\Big),
\end{equation}
where $P$ and the delta-function denote the Cauchy principal value
and the singularity, respectively. The integro-differential equation
in (\ref{t13}) is thus reduced to a linear ordinary differential
equation. The solutions of $b_{0}^{\prime }$ as well as $b_{0}$ can
then be easily obtained as $b_{0}(t)=e^{-i(\omega _{0}-\delta \omega
)t-\pi J(\omega _{0})t} $, where $\delta \omega =P\int_{0}^{\infty
}\frac{J(\omega )d\omega }{\omega -\omega _{0}}$. Thus one can
verify that,
\begin{equation}
\gamma (t)\equiv \gamma _{0}=2\pi J(\omega _{0}),~~\Omega (t)\equiv
\Omega _{0}=2(\omega _{0}-\delta \omega ),  \label{t18}
\end{equation}%
which are just the coefficients in the Markovian master equation of
the two-level atom system \cite{Breuer02}.

\section{Purity and decoherence factor}\label{quant}

To quantify the decoherence dynamics of the qubit, we introduce the
following two quantities. The first one is the purity, which is
defined as \cite{Nielsen00}
\begin{equation}
p(t)=Tr\rho ^{2}(t).  \label{t7}
\end{equation}%
Clearly $p=1$ for pure state and $p<1$ for mixed state. The second
quantity describing the decoherence is the decoherence factor $c(t)$
of the qubit, which is determined by the off-diagonal elements of
the reduced density matrix
\begin{equation}
\left\vert \rho _{12}(t)\right\vert =c(t)\left\vert \rho
_{12}(0)\right\vert .  \label{t8}
\end{equation}
The decoherence factor maintains unity when the reservoir is absent
and vanishes for the case of completely decoherence.

For definiteness, we consider the following initial pure state of
the qubit
\begin{equation}
\left\vert \psi (0)\right\rangle =\alpha \left\vert +\right\rangle
+\beta \left\vert -\right\rangle ,  \label{t9}
\end{equation}%
in which $\alpha $ and $\beta $ satisfy the normalization condition.
Using Eq. (\ref{t5}), the exact time evolution of the qubit is
easily
obtained%
\begin{equation}
\rho (t)=\left(
\begin{array}{cc}
\left\vert \alpha \right\vert ^{2}\left\vert b_{0}(t)\right\vert
^{2} &
\alpha \beta ^{\ast }b_{0}(t) \\
\alpha ^{\ast }\beta b_{0}^{\ast }(t) & 1-\left\vert \alpha
\right\vert
^{2}\left\vert b_{0}(t)\right\vert ^{2}%
\end{array}
\right) .  \label{t10}
\end{equation}
With Eq. (\ref{t10}), the purity and decoherence factor can be
expressed explicitly
\begin{equation}
p(t)=2\left\vert \alpha \right\vert ^{4}\left\vert
b_{0}(t)\right\vert ^{2}[\left\vert b_{0}(t)\right\vert ^{2}-1]+1,
\label{t11}
\end{equation}
and
\begin{equation}
c(t)=\left\vert b_{0}(t)\right\vert .  \label{t12}
\end{equation}

It is easy to verify, under the Born-Markovian approximation, the
purity and
decoherence factor have the following forms%
\begin{equation}
p(t)=2\left\vert \alpha \right\vert ^{4}e^{-\gamma _{0}t}(e^{-\gamma
_{0}t}-1)+1,  \label{t16}
\end{equation}
and
\begin{equation}
c(t)=e^{-\frac{\gamma _{0}}{2}t},  \label{t17}
\end{equation}
where the time-independent decay rate $\gamma_0$ is given in Eq.
(\ref{t18}). Obviously, the system asymptotically loses its quantum
coherence ($c(\infty)=0$) and approaches a pure steady state
($p(\infty)=1$) irrespective of the form of the initial state under
the Markovian approximation. One can also find from Eqs.
(\ref{t11}-\ref{t17}) that the probability amplitude of excited
state plays key role in the decoherence dynamics.

\section{Numerical results and analysis}\label{numa}
\begin{figure*}[tbp]
\centering
\includegraphics[width = \columnwidth]{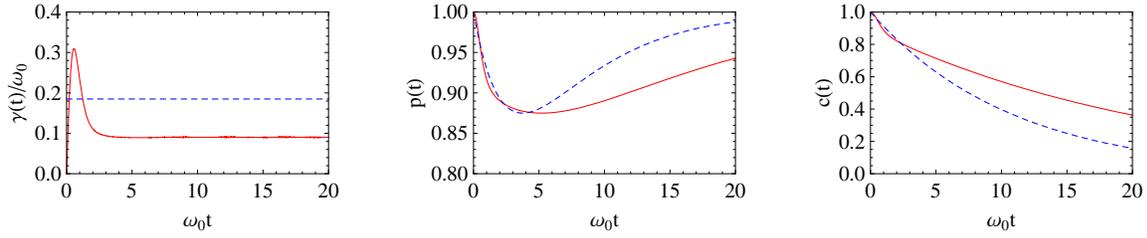}
\caption{(Color online). Time evolution of $\gamma (t)$, $p(t)$ and
$c(t)$ in non-Markovian situation (solid line) and the corresponding
Markovian situation (dashed line), when $\eta$ and $\omega_c$ are
small. The parameters used here are $\alpha =1/\sqrt{2}$, $\eta
=0.08$ and $\omega _{c}/\omega _{0}=1.0.$}\label{weakf}
\end{figure*}

In this section, by numerically solving Eq. (\ref{t4}), we study the
influence of memory effect of reservoir on the exact dynamics of the
qubit. Noticing the fact that the memory effect registered in the
kernel function is essentially determined by the spectrum density
$J(\omega )$, one can expect that $J(\omega)$ plays an major role in
the exact dynamics of the qubit. In the following, we show how the
decoherence of the qubit can be fully suppressed under the
non-Markovian dynamics in terms of the relevant parameters of
$J(\omega)$.

\subsection{The influence of coupling constant}

In the following, we numerically analyze the exact decoherence
dynamics of the qubit with respect to decay rate $\gamma (t)$,
purity $p(t)$ and decoherence factor $c(t)$ in terms of the coupling
constant $\eta $.

In Fig. \ref{weakf} we plot the time evolution of decay rate
$\gamma(t)$, purity $p(t)$, decoherence factor $c(t)$ and their
Markovian correspondences in the weak coupling and low cutoff
frequency case. We can see that $\gamma(t)$ shows distinct
difference from its Markovian counterpart over a very short time
interval. With time, $\gamma(t)$ tends to a definite positive value.
The small ``jolt" of $\gamma(t)$ in the short time interval just
evidences the backaction of the memory effect of the reservoir
exerted on the qubit \cite{Hu92,An07}. It manifests that the reservoir
does not exert decoherence on the qubit abruptly, just as the result
based on Markovian approximation, but dynamically influences the
qubit and gradually establishes a stable decay rate to the qubit.
Furthermore, it is also shown that the decay rate is positive in the
full range of evolution, which results in any initial qubit state
evolving to the ground state $\left\vert \psi (\infty)\right\rangle
=\left\vert -\right\rangle $ irreversibly. Consequently the
decoherence factor monotonously decreases to zero with time and the
purity approaches unity in the long-time limit, which is consistent
with the result under Markovian approximation. The result indicates
that although the reservoir has backaction effect on the qubit, it
is quite small. And the dissipation effect of the reservoir
dominates the dynamics of the qubit. Thus no qualitative difference
can be expected between the exact result and the Markovian one with
the backaction effect ignored. Therefore the widely used Markovian
approximation is applicable in this case. Nevertheless, at the short
and immediate time scales the overall behavior is still quite
different from that of the Markovian dynamics. The decoherence
factor shown in the righ-hand panel of Fig. \ref{weakf} shows
non-exponential decay, which is in agreement with the result
obtained previously in the spin-boson model in the weak-coupling
limit \cite{Loss05}. However, the situation is dramatically changed
if the coupling is strengthened as discussed below.
\begin{figure*}[tbp]
\centering
\includegraphics[width = \columnwidth]{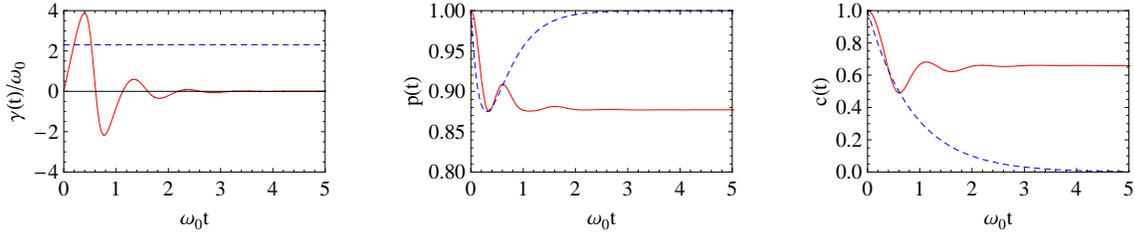}
\caption{(Color online). Time evolution of $\gamma (t)$, $p(t)$ and
$c(t)$ in non-Markovian situation (solid line) and the corresponding
Markovian situation(dashed line), when $\eta$ is large. The
parameters used here are $\alpha =1/\sqrt{2}$, $\eta =1.0$ and
$\omega _{c}/\omega _{0}=1.0.$}\label{strongf}
\end{figure*}

With the same cutoff frequency as in Fig. \ref{weakf} but a larger
coupling constant, we plot in Fig. \ref{strongf} the decay rate,
purity and decoherence factor in the strong coupling case. In this
case the non-negligible backaction of the reservoir has a great
impact on the dynamics of the qubit. Firstly, we can see that the
decay rate not only exhibits oscillations, but also takes negative
values in the short time scale. Physically, the negative decay rate
is a sign of strong backaction induced by the non-Markovian memory
effect of the reservoir. And the oscillations of the decay rate
between negative and positive values reflect the exchange of
excitation back and forth between qubit and the reservoir
\cite{Pillo08}. Consequently both the decoherence factor and the
purity exhibit oscillations in a short-time scale, which shows
dramatic deviation to the Markovian result. Therefore, entirely
different to the weak coupling case in Fig. \ref{weakf}, the
reservoir in the strong coupling case here has strong backaction
effect on the qubit. Secondly, we also notice that the decay rate
approaches zero in the long-time limit. The vanishing decay rate
means, after several rounds of oscillation, the qubit ceases
decaying asymptotically. The non-Markovian purity maintains a steady
value asymptotically, which is less then unity. This indicates that
the steady state of the qubit is not the ground state anymore, but a
mixed state. The decoherence factor also tends to a non-zero value,
which implies that the coherence of the qubit is preserved with a
noticeable fraction in the long-time steady state. These phenomena,
which are qualitatively different to the Markovian situation,
manifest that the memory effect has a considerable contribution not
only to the short-time, but also to the long-time behavior of the
decoherence dynamics. The presence of the residual coherence in the
steady state also suggests a potential active control way to protect
quantum coherence of the qubit from decoherence via the
non-Markovian effect.
\begin{figure*}[tbp]
\centering
\includegraphics[width = \columnwidth]{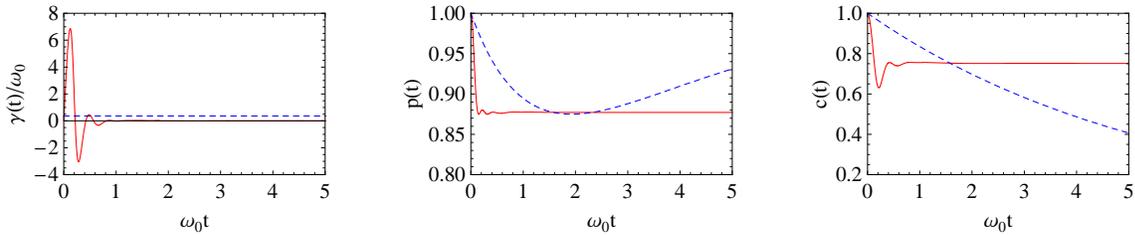}
\caption{(Color online). Time evolution of $\gamma (t)$, $p(t)$ and
$c(t)$ in non-Markovian situation(solid line)and the corresponding
Markovian situation(dashed line), when $\omega _{c}$
is large. The parameters used here are $\alpha =1/\sqrt{2}$%
, $\eta =0.08$ and $\omega _{c}/\omega _{0}=3.0.$}\label{sc}
\end{figure*}

\subsection{The influence of cutoff frequency}

The cutoff frequency $\omega _{c}$, on the one hand, is introduced
to eliminate the infinity in the frequency integration. On the other
hand it also determines the frequency range in which the power form
is valid \cite{Weiss}. In the following, we elucidate the influence
of cutoff frequency on the exact decoherence dynamics.

Fixing $\eta$ as the value in Fig. \ref{weakf} and increasing the
cutoff frequency, we plot in Fig. \ref{sc} the dynamics of the qubit
in a high cutoff frequency case. It shows that a similar decoherence
behavior as the strong coupling case in Fig. \ref{strongf} can be
obtained. After several rounds of oscillation, the decay rate tends
to zero in the long-time limit. The negative decay rate makes the
lost coherence partially recovered. The vanishing decay rate in the
long-time limit results in the decoherence frozen before the qubit
gets to its ground state. Thus there is some residual coherence
trapped in the steady state. Similar to the strong coupling case, it
is essentially the interplay between the backaction and the
dissipation on the dynamics of qubit which results in the inhibition
of decoherence. We argue that in this high cutoff frequency regime,
the widely used Markovian approximation is not applicable because of
the strong backaction effect of the reservoir.

\subsection{The physical mechanism of the decoherence
inhibition}From the analysis above we can see clearly that the
decoherence can be inhibited in the non-Markovian dynamics. A
natural question is: What is physical mechanism to cause such
dynamical decoherence inhibition? To answer this question, let us
find the eigen solution of Eq. (\ref{t1}) in the sector of
one-excitation in which we are interested. The eigenequation reads
$H\left\vert \varphi _{E}\right\rangle =E\left\vert \varphi
_{E}\right\rangle $, where $\left\vert \varphi _{E}\right\rangle
=c_{0}\left\vert +,\{0_{k}\}\right\rangle +\sum_{k=0}^{\infty
}c_{k}\left\vert -,1_{k}\right\rangle $. After some algebraic
calculation, we can obtain a transcendental equation of $E$
\begin{equation}
y(E)\equiv\omega _{0}- \int_{0}^{\infty }\frac{J(\omega) }{\omega
-E}d\omega =E. \label{eigen}
\end{equation}
From the fact that $y(E)$ decreases monotonically with the increase
of $E$ when $E<0$ we can say that if the condition $y(0)<0$, i.e.
\begin{equation} \omega _{0}-2\eta
\frac{\omega^3 _{c}}{\omega_0^2}<0\label{cd}\end{equation} is satisfied, $y(E)$ always has one and only one intersection in the regime $%
E<0$ with the function on the right-hand side of Eq. (\ref{eigen}).
Then the system will have an eigenstate with real (negative)
eigenvalue, which is a bound state \cite{Miyamoto}, in the Hilbert
space of the qubit plus its reservoir. While in the regime of $E>0$,
one can see that $y(E)$ is divergent, which means that no real root
$E$ can make Eq. (\ref{eigen}) well-defined. Consequently Eq.
(\ref{eigen}) does not have positive real root to support the
existence of a further bound state. It is noted that Eq.
(\ref{eigen}) may possess complex root. Physically this means that
the corresponding eigenstate experiences decay contributed from the
imaginary part of the eigenvalue during the time evolution, which
causes the excited-state population approaching zero asymptotically
and the decoherence of the reduced qubit system.

The formation of bound state is just the physical mechanism
responsible for the inhibition of decoherence. This is because a
bound state is actually a stationary state with a vanishing decay
rate during the time evolution. Thus the population probability of
the atomic excited state in bound state is constant in time, which
is named as ``population trapping"
\cite{Yablonovitch87,Lambropoulos00}. This claim is fully verified
by our numerical results. The parameters in Fig. \ref{weakf} do not
satisfy the condition (\ref{cd}) to support the existence of a bound
state, then the dynamics experiences a severe decoherence. While
with the increase of either $\eta$ (in Fig. \ref{strongf}) or
$\omega_c$ (in Fig. \ref{sc}), the bound state is formed. Then the
system and its environment is so correlated that it causes the decay
rate of the system in the non-Markovian dynamics exhibiting: 1)
transient negative value due to the backaction of the environment;
2) vanishing asymptotic value. Such interesting phenomenon, i.e. the
vanishing asymptotical decay rate in the large cutoff frequency
regime for super-Ohmic spectrum density, was also revealed in Ref.
\cite{Paz09}. This effect of course is missing in the conventional
Born-Markovian decoherence theory, where the reservoir is
memoryless.

\begin{figure}[tbp]
\centering
\includegraphics[scale=0.65]{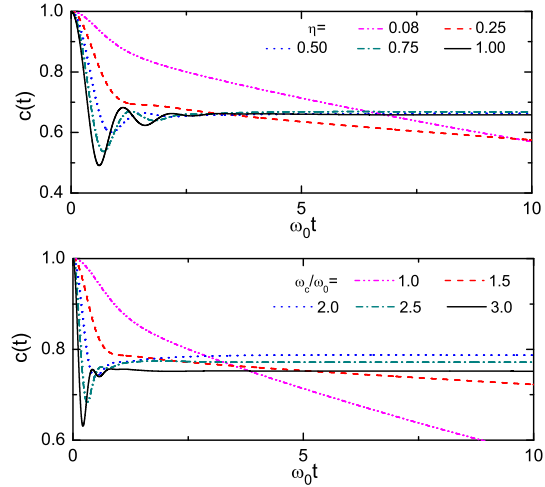}
\caption{(Color online). Time evolution of $c(t)$ in the
non-Markovian dynamics with different $\eta$ when
$\omega_c/\omega_0=1.0 $ (upper panel) and with different $\omega_c$
when $\eta=0.08$ (lower panel).}\label{comp}
\end{figure}

In order to understand the exact decoherence dynamics more
completely, we plot in Fig. \ref{comp} the crossover from coherence
destroying to coherence trapping via increasing either the coupling
constant or the cutoff frequency. Coherence trapping can be achieved
as long as the bound state is formed. Therefore, one can preserve
coherence via tuning the relevant parameters of system and the
reservoir, e.g. the qubit-reservoir coupling constant and the
property of the reservoir so that the condition (\ref{cd}) is
satisfied.

\section{Summary and discussions}\label{dsumma}
In summary, we have investigated the exact decoherence dynamics of a
qubit in a dissipative vacuum reservoir. We have found that even in
a vacuum environment without any nontrivial structure, we can still
get the decoherence suppression of the qubit owing to the dynamical
mechanism of the non-Markovian effect. From our analytic and
numerical results, we find that the non-Markovian reservoir has dual
effects on the qubit: dissipation and backaction. The dissipation
effect exhausts the coherence of the qubit, whereas the backaction
one revives it. In the strong coupling and/or high cutoff frequency
regimes, a bound state between the qubit and its reservoir is
formed. It induces a strong backaction effect in the dynamics
because the reservoir is strongly correlated with the qubit in the
bound state. Furthermore, because of the non-decay character of the
bound state the decay rate in this situation approach zero
asymptotically. The vanishing of the decay rate causes the
decoherence to cease before the qubit decays to its ground state.
Thus the qubit in the non-Markovian dynamics would evolve to a
non-ground steady state and there is some residual coherence
preserved in the long-time limit. Our results make it clear how the
non-Markovian effect shows its effects on the decoherence dynamics
in different parameter regimes.

The presence of such coherence trapping phenomenon actually gives us
an active way to suppress decoherence via non-Markovian effect. This
could be achieved by modifying the properties of the reservoir to
approach the non-Markovian regime via the potential usage of the
reservoir engineering technique
\cite{Wineland001,Wineland002,Diehl08,Garraway}. Many experimental
platforms, e.g. mesoscopic ion trap \cite{Wineland001,Wineland002},
cold atom BEC \cite{Diehl08}, and the photonic crystal material
\cite{Yablonovitch87} have exhibited the controllability of
decoherence behavior of relevant quantum system via well designing
the size (i.e. modifying the spectrum) of the reservoir and/or the
coupling strength between the system and the reservoir. It is also
worth mentioning that a proposal aimed at simulating the spin-Boson
model, which is relevant to the one considered in this paper, has
been reported in the trapped ion system \cite{Porras}. On the other
side many practical systems can now be engineered to show the novel
non-Markovian effect \cite{Dubin07,Koppens07,Mogilevtsev08,Xu09}.
All these achievements show that the recent advances have paved the
way to experimentally simulate the paradigmatic models of open
quantum system, which is one part of the new-emergent field, quantum
simulators \cite{Nori}. Our work sheds new light on the way to
indirectly control and manipulate the dynamics of quantum system in
this experimental platforms.

A final remark is that our results can be generalized to the system
consisted of two qubits, each of which interacts with a local
reservoir. Because of the coherence trapping we expect that the
non-Markovian effect plays constructive role in the entanglement
preservation \cite{Bellomo08,An09,Tong09}.

\section*{Acknowledgments} This work is supported by NSF of China under Grant No. 10604025,
Gansu Provincial NSF of China under Grant No. 0803RJZA095, Program
for NCET, and NUS Research Grant No. R-144-000-189-305.

\end{document}